\documentclass[a4paper,11pt]{article}

\usepackage{lmodern}
 \usepackage{tabularx}
 \usepackage{amsmath,amsthm,amssymb}
 \usepackage{makeidx,epsfig,lscape}
 \usepackage{array}
 \usepackage[most]{tcolorbox}
 \usepackage{wrapfig, framed, caption}
 \usepackage{mathtools}
 \usepackage{blindtext}
 \usepackage{siunitx}
 \usepackage{tikz}
 \usepackage{multirow}
 \usepackage{color,colortbl}
 \usepackage{fancyhdr}
 \usepackage{xcolor,pict2e}
\usepackage{mdframed}
\usepackage{epsfig}
\usepackage{lineno}
\usepackage{physics}
\usepackage{stackengine}
\usepackage{scalerel}
\usepackage{esvect}
\usepackage{enumitem}
\usepackage{siunitx}
\usepackage[document]{ragged2e}
\usepackage[export]{adjustbox}
\usepackage[font={small}]{caption}
\usepackage{indentfirst}
\renewcommand{\headrulewidth}{0pt}
\renewcommand{\footrulewidth}{0.5pt}
\definecolor{myaqua}{rgb}{0.0,0.5,0.55}
\definecolor{lightaqua}{rgb}{0.75,0.95,0.95}
\definecolor{maroon}{rgb}{0.824,0.137,0.169}

\usepackage[colorlinks = true,
linkcolor = myaqua,
urlcolor  = blue,
citecolor = myaqua]{hyperref}

\usepackage{caption}
\usepackage{floatrow}

\newcolumntype{C}[1]{>{\centering\let\newline\\\arraybackslash\hspace{0pt}}m{#1}}
\newcolumntype{L}[1]{>{\raggedright\let\newline\\\arraybackslash\hspace{0pt}}m{#1}}

\newcommand{\vectarrow}[1]{\vec{\boldsymbol{#1}}} 

\newcommand{\xhat}{\hat{\textit{$\boldsymbol{x}$}}}
\newcommand{\yhat}{\hat{\textit{$\boldsymbol{y}$}}}
\newcommand{\zhat}{\hat{\textit{$\boldsymbol{z}$}}}

\definecolor{nblue}{rgb}{0, 0, 1}
\definecolor{nred}{rgb}{0.97, 0.02, 0}
\definecolor{ngreen}{rgb}{0.06, 0.40, 0.05}
\definecolor{byzantine}{rgb}{0.74, 0.2, 0.64}
\definecolor{cerisepink}{rgb}{0.984, 0.45, 0.97}
\let\textquotedbl="

\def\lin#1#2{\textcolor[rgb]{0.6,0.6,0.6}{\vspace*{#1mm} \hrule
		height 3 pt \vspace*{#2mm}}}
	
\def\bt{\begin{tabular}}
	\def\et{\end{tabular}}
\def\and{\mbox{ and }}

\def\1{{\bf 1}}

\def\sectionn#1{\refstepcounter{section}{\color{maroon}		
\vskip 6mm		
\noindent\Large\bf\thesection. #1}	
\vskip 3mm}

\begin{document}

 \fancyhead[L]{\hspace*{-13mm}
 \bt{l}{\bf Open Journal of *****, 2019, *,**}\\
 Published Online **** 2019 in SciRes.
 \href{http://www.scirp.org/journal/*****}{\color{blue}{\underline{\smash{http://www.scirp.org/journal/****}}}} \\
 \href{http://dx.doi.org/10.4236/****.2019.*****}{\color{blue}{\underline{\smash{http://dx.doi.org/10.4236/****.2019.*****}}}} \\
 \et}

 $\mbox{ }$
{{\noindent{\huge Amplification of polarization correlations in Compton scattering of hard X-ray Bell states}}\\[6mm]

{ \small\bf Peter Caradonna} } 
\\
{Kavli Institute for the Physics and Mathematics of the Universe (WPI), Institutes for Advanced Study (UTIAS), The University of Tokyo,
5-1-5 Kashiwa-no-Ha, Kashiwa, Chiba, 277-8583, Japan}\\

\noindent{\textbf{Email}: \href{mailto:pietro.caradonna@ipmu.jp}{\color{blue}{\smash{pietro.caradonna@ipmu.jp}}}}\\[4mm]
{\noindent{\bf Keywords:}{~Entanglement, Compton scattering, Bell states, hard X-rays, Incoherent scattering cross sections}
 \lin{5}{7}

{\noindent{\large\textbf{Abstract}}{\\
\fontfamily{times}\normalsize\selectfont\textup{The theoretical cross section for Compton scattering of maximally entangled Bell photons has yet to be rigorously confirmed by experiments. Test cases of Bell states for use in Compton scattering experiments can now be expanded given reports of creating all 4 Bell states in the hard X-ray regime by the process of spontaneous parametric down-conversion. We outline an experiment and apply a matrix method to parameterize Compton scattering theory using the phase matching angles. When the azimuths of two hypothetical photon counters are recorded at angles of 0 degrees and 90 degrees, and the ratio of their counting rates determined, azimuthal ratios are expected to be 600 times larger compared to 511 keV Bell photons.}}}
\\
\lin{1}{1}  
\renewcommand{\headrulewidth}{0.5pt}
\renewcommand{\footrulewidth}{0.5pt}
 
 \pagestyle{fancy}
 \fancyfoot{}
 \fancyhead{} 
 \fancyhf{}
 \fancyhead[RO]{\leavevmode \put(-90,0){P. Caradonna \textit{et al.}}}
 \fancyhead[LE]{\leavevmode \put(0,0){P. Caradonna \textit{et al.}}}
 \fancyfoot[C]{\leavevmode
 \put(-2.5,-30){\thepage}}
 
 \renewcommand{\headrule}{\hbox to\headwidth{\leaders\hrule height \headrulewidth\hfill}}
 
\mdfdefinestyle{mystyle}{backgroundcolor=white!}

\sectionn{Introduction}
{ \fontfamily{times}\selectfont
Simulations show that image quality in Positron Emission Tomography can be improved by exploiting the polarization entanglement of 511 keV photons~\cite{McNamara2014}. These entangled states were also used in Compton scattering experiments~\cite{Kasday1971,Kasday1975} to test the kind of entanglement discussed by Einstein, Podolsky and Rosen~\cite{Einstein1935,Bohm1957,Bohm1976}. However, these experiments had to assume theory can predict the angular distributions to a similar accuracy as for non-entangled photons~\cite{Heitler1954}. To this day, this assumption must be invoked because the question concerning the accuracy of Compton scattering theory in the presence of entanglement still remains open~\cite{Caradonna2019, Hiesmayr2019}.  

The main contribution of this work is to provide predictions for experiments concerned with closing the question regarding the accuracy of Compton scattering theory for Bell state photons. To date, only the 511 keV cross-polarized Bell state has been available for Compton scattering experiments. Although much research has been conducted using this system, the most accurate measurement of a quantity known as the `azimuthal ratio' yielded maximum ratio values that are $87\%$ or less than the predicted value~\cite{Kasday1975,Caradonna2019,Langhoff1960}. Development of novel X-ray sources~\cite{Bernhard2013,Yabashi2001} which can be used to generate any one of the 4 Bell states in the hard X-ray energy region~\cite{Shwartz2011} is appealing, because azimuthal ratios can now, in principle, be measured using controllable incident beams rather than relying solely on positron emitting sources~\cite{Kasday1971,Kasday1975,Langhoff1960}. 

A fundamental difference between non-entangled photons and Bell state photons is that the former are either polarized {\it{or}} unpolarized, whereas the latter are both polarized {\it{and}} unpolarized. This can be illustrated by first considering the 4 Bell photon state wavefunctions in the linearly polarization basis given by  
\begin{subequations}
	\begin{equation}
	\ket{\Psi^{\pm}_{l}} = \frac{1}{\sqrt{2}}
	\begin{bmatrix}	1 \\ 0 \end{bmatrix}\otimes\begin{bmatrix}	0 \\ 1 \end{bmatrix}\pm\frac{1}{\sqrt{2}}	\begin{bmatrix}	0 \\ 1 \end{bmatrix}\otimes\begin{bmatrix}	1 \\ 0 \end{bmatrix},
	\label{eqn:1a}
	\end{equation}
	\begin{equation}
	\ket{\Phi^{\pm}_{l}} = \frac{1}{\sqrt{2}}
	\begin{bmatrix}	1 \\ 0 \end{bmatrix}\otimes\begin{bmatrix}	1 \\ 0 \end{bmatrix}\pm\frac{1}{\sqrt{2}}	\begin{bmatrix}	0 \\ 1 \end{bmatrix}\otimes\begin{bmatrix}	0 \\ 1 \end{bmatrix},
	\label{eqn:1b}
	\end{equation}
	\label{eqn:1}
\end{subequations}
where the cross-polarization entangled states $\ket{\Psi^{+}_{l}}$ and $\ket{\Psi^{-}_{l}}$ only differ by a phase factor, and likewise for the parallel-polarization entangled states $\ket{\Phi^{+}_{l}}$ and $\ket{\Phi^{-}_{l}}$. It will be shown in Sec. (\ref{sec:Section2}) that a source of Bell state photons will consist of two photon beams each of which can travel along different trajectories. Each beam in of itself is unpolarized. This dichotomy of Bell photons being both polarized (with respect to each other) and unpolarized (with respect to space-like separated photon detectors) manifests itself in the double differential Compton collision cross section. For example, the cross section for the states $\ket{\Psi^{\pm}_{l}}$ can be expressed as follows~\cite{Caradonna2019}
\begin{equation}
\frac{\partial^{2}\sigma_{KN}^{(\Psi_{l}^{\pm})}}{\partial\Omega_{1}\partial\Omega_{2}}=\frac{1}{2}\frac{d\sigma_{KN}}{d\Omega_{1}}\frac{d\sigma_{KN}}{d\Omega_{2}}-\frac{\partial^{2}\sigma_{KN}^{(\Phi_{l}^{\pm})}}{\partial\Omega_{1}\partial\Omega_{2}}.
\label{eqn:2}
\end{equation}
As can be seen, this cross section can be described as the difference of two terms: The term $\left(d\sigma_{KN}/d\Omega_{1}\right)\left(d\sigma_{KN}/d\Omega_{2}\right)$ represents the product of Klein-Nishina (KN) cross sections describing Compton scattering by a pair of mutually independent and unpolarized photons. The second term represents the cross section for parallel-polarization entangled states $\ket{\Phi_{l}^{\pm}}$.

\sectionn{Production of hard X-ray Bell states}
\label{sec:Section2}
Figure (\ref{fig:Fig1}) is a sketch of a parametric down-conversion experiment, which assumes diamond as the nonlinear medium for creating Bell states photons. The polarization correlations are analyzed down stream using scattering and photo-absorption based mediums. 

Figure (\ref{fig:Fig1}) depicts entangled photons created by a pump photon with wave vector $\vectarrow{k}_{p}$ directed towards and impacting on a diamond target. In this example, a pair of maximally entangled signal and idler photon are emitted with wave vectors $\vectarrow{k}_{s}$ and $\vectarrow{k}_{i}$, respectively. The vectors $\vectarrow{k}_{p}$, $\vectarrow{k}_{s}$ and $\vectarrow{k}_{i}$ subtend angles of $\vartheta_{p}$, $\vartheta_{s}$ and $\vartheta_{i}$, respectively, with respect to the (111) atomic planes. The pump photon is assumed to be linearly polarized normal to a {\it{`trajectory plane'}} defined by the three vectors $\vectarrow{k}_{p}$, $\vectarrow{k}_{s}$ and $\vectarrow{k}_{i}$. To generate Bell state photons, these vectors must satisfy the vector condition $\vectarrow{k}_{s}+\vectarrow{k}_{i}=\vectarrow{k}_{p}+\vectarrow{G}$, where $\vectarrow{G}$ denotes the reciprocal lattice vector. The vector diagram shown near the bottom right of Fig. (\ref{fig:Fig1}) visualizes this vector condition. 

Figure (\ref{fig:Fig1}) visualizes the creation of cross-polarized entangled signal and idler photons described by Eqn. (\ref{eqn:1a}) wavefunctions. Attached to each photon are two system of coordinates. The first set of coordinate systems are labeled $\{x_{s},$ $ y_{s},$ $ z_{s}\}$ and $\{x_{i}, y_{i}, z_{i}\}$, where the subscript $s$ and $i$ label the coordinate system of the `signal' and `idler', respectively. The $\{y_{s},$ $ z_{s}\}$ and $\{y_{i},$ $ z_{i}\}$ axis lie in the trajectory plane and the $x_{s}$ and 
$x_{i}$ axis are perpendicular to it. The photons themselves are represented as wave packets that travel along their respective wave vectors. Prior to a measurement, these photons are in a quantum superposition of cross-polarized states. This is approximated in the sketch by a pair of photons in a superposition of blue and green wave packets, which is
\begin{framed}
	\begin{center}
		\includegraphics[width=\linewidth,frame]{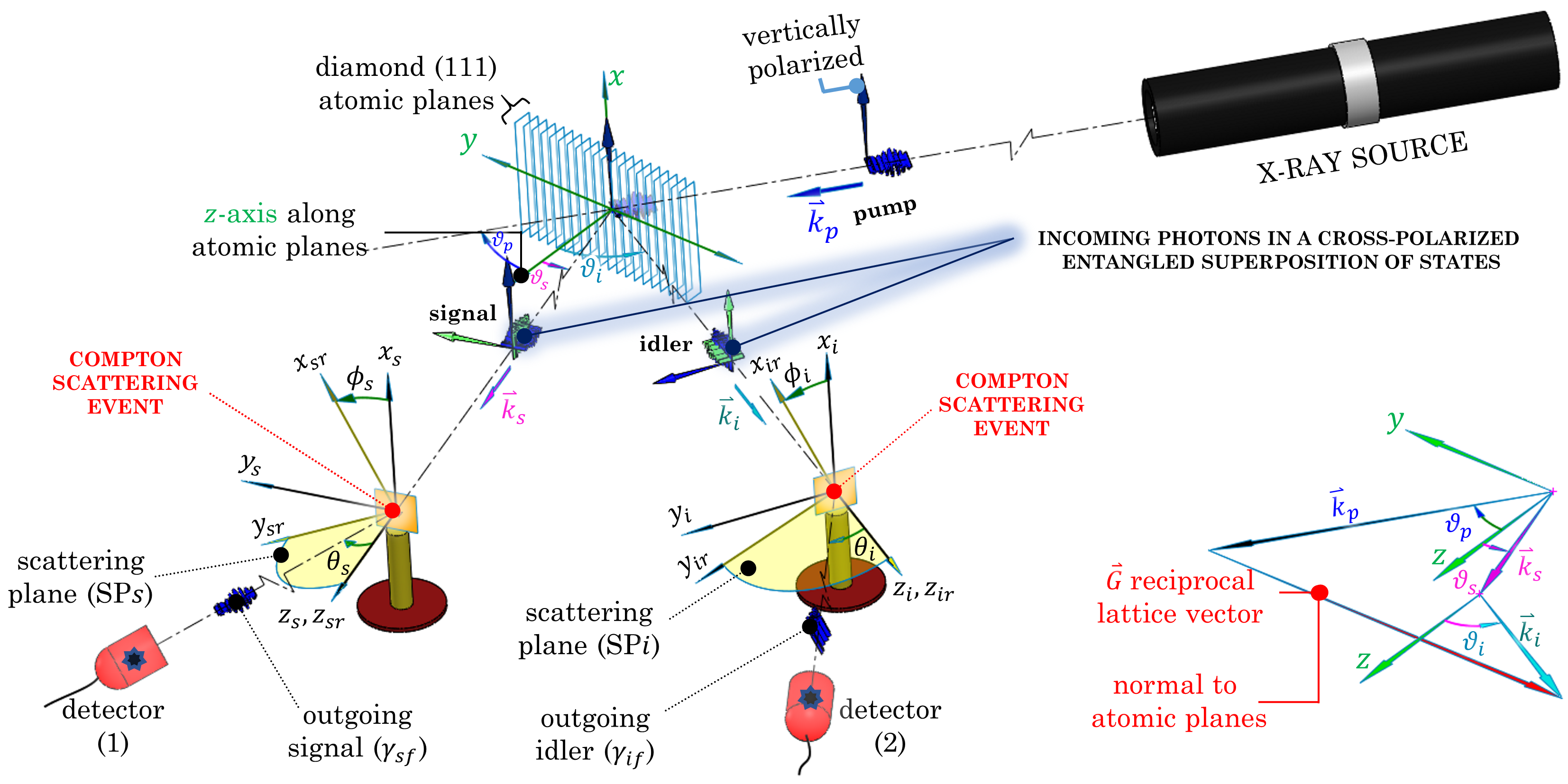}
		\captionof{figure}{Schematic of an experimental setup where the wave vectors $\vectarrow{k}_{i}$, $\vectarrow{k}_{s}$, and $\vectarrow{k}_{p}$ label the wave vectors of the idler, signal, and pump photons, respectively, and these vectors lie in a plane which we define as the `trajectory plane'.} 
		\label{fig:Fig1}
	\end{center}
\end{framed}
to mean that the Bell state wavefunction will collapse to either a pair of blue or green photon states in any given Compton scattering coincidence measurement. The linear polarization state of each photon is represented by arrows which point in the direction of polarization. With respect to their coordinate systems, these arrows point either along the $\xhat$ or $\yhat$ directions and represent a vertical and horizontal polarization state, respectively.

The second set of coordinates systems are labeled $\{x_{sr}, $ $y_{sr},$ $ z_{sr}\}$ and $\{x_{ir},$ $ y_{ir},$ $ z_{ir}\}$, where the subscript `$r$' denotes `rotated'. With respect to the first coordinate system, we have the condition $\zhat_{sr}\cdot\zhat_{s} = \zhat_{ir}\cdot\zhat_{i} = 1$, $\xhat_{sr}\cdot\xhat_{s} = \cos\phi_{s}$, and $\xhat_{ir}\cdot\xhat_{i} = \cos\phi_{i}$, where $\phi$ denotes the azimuthal angle. The orientation of these rotated system of coordinates are defined only by the Compton scattering process itself, since the Compton scattering planes of the signal and idler photons, labelled SP{\it{s}} and SP{\it{i}}, are defined by the $\{y_{sr},$ $ z_{sr}\}$ and $\{y_{ir},$ $ z_{ir}\}$ axis, respectively. The angles $\theta_{i}$ and $\theta_{s}$ are the Compton angles which subtend the incident and scattered wave vectors. Throughout this manuscript we have used the cross-polarized entangled states as a visualization aid to motivate the theoretical analysis of all 4 Bell states given in Eqn. (\ref{eqn:1}).
\sectionn{Departures from the Klein-Nishina model}
\label{sec:Section3}
The cross section given in Eqn. (\ref{eqn:2}) is based on the Klein-Nishina model which assumes photons Compton scatter off stationary electrons in a zero kelvin temperature environment. Departures from this model occur in situations where the photon incident energy $E_{o}$ is comparable with the binding energy $E_{b}$ of the inner-shell electron of the target. The incoherent scattering cross sections $\partial^{2}\sigma_{incoh}/\partial\Omega_{i}\partial\Omega_{s}$ for states $\ket{\Psi_{l}^{\pm}}$ and $\ket{\Phi_{l}^{\pm}}$ now replaces the Klein-Nishina model in a heuristic way~\cite{Ribberfors1982} such that
\begin{subequations}
	\begin{equation}
		\begin{aligned}
			\frac{\partial^{2}\sigma_{incoh}^{\left(\Psi_{l}^{\pm}\right)}}{\partial\Omega_{i}\partial\Omega_{s}}=\frac{\partial^{2}\sigma_{KN}^{\left(\Psi_{l}^{\pm}\right)}}{\partial\Omega_{i}\partial\Omega_{s}}S(\theta_{i},\theta_{s};E_{oi},E_{os},Z_{i},Z_{s}),
			\label{eqn:3a}
		\end{aligned}
	\end{equation}
	and	
	\begin{equation}
		\begin{aligned}
			\frac{\partial^{2}\sigma_{incoh}^{\left(\Phi_{l}^{\pm}\right)}}{\partial\Omega_{i}\partial\Omega_{s}}=\frac{\partial^{2}\sigma_{KN}^{\left(\Phi_{l}^{\pm}\right)}}{\partial\Omega_{i}\partial\Omega_{s}}S(\theta_{i},\theta_{s};E_{oi},E_{os},Z_{i},Z_{s}),
		\end{aligned}
		\label{eqn:3b}
	\end{equation}
	\label{eqn:3}
\end{subequations}
where it has been shown by Ribberfors and Berggren that the factorization on the right of the equal sign appears naturally out of the direct integration of the relativistic formulation of the Compton profiles~\cite{Ribberfors1982}, and where 
\begin{subequations}
	\begin{equation}
		\begin{aligned}
			&\frac{\partial^{2}\sigma_{KN}^{\left(\Psi_{l}^{\pm}\right)}}{\partial\Omega_{i}\partial\Omega_{s}}=\frac{1}{8}\sum\limits_{a=0}^{1}\bra{I}T\left(\theta_{i};E_{oi}\right)M\left(\frac{a\pi}{2}\right)\ket{+}\bra{I}T\left(\theta_{s};E_{os}\right)M\left(\eta+\frac{a\pi}{2}\right)\ket{-},
			\label{eqn:4a}
		\end{aligned}
	\end{equation}
	
	\begin{equation}
		\begin{aligned}
			&\frac{\partial^{2}\sigma_{KN}^{\left(\Phi_{l}^{\pm}\right)}}{\partial\Omega_{i}\partial\Omega_{s}}=\frac{1}{8}\sum\limits_{a=0}^{1}\bra{I}T\left(\theta_{i};E_{oi}\right)M\left(\frac{a\pi}{2}\right)\ket{+}\bra{I}T\left(\theta_{s};E_{os}\right)M\left(\eta+\frac{a\pi}{2}\right)\ket{+},
		\end{aligned}
		\label{eqn:4b}
	\end{equation}
	\label{eqn:4}
\end{subequations}
are the matrix representation of the cross sections derived using the pioneering work of Wightman~\cite{Wightman1948}, Fano~\cite{Fano1949}, and McMaster~\cite{McMaster1961}, and the reader is referred to~\cite{Caradonna2019} and references therein for more detail. The cross sections have been averaged and summed over the initial and final electron spin states and summed over the final photon polarization states. The Stokes vectors $\ket{+}$ and $\ket{-}$ represent a vertical and horizontal polarization state, respectively. The incident photon energies $E_{oi}$ and $E_{os}$ are given in units of $mc^{2}$, the bra vector $\bra{I}$ represents an insensitive-to-polarization photon counter, $T$ is the Compton transition matrix, $M$ is a matrix which transforms the Stokes vectors from the first to the second system of coordinates, and $\eta$ is the relative angle between the two scattering planes such that
\begin{equation}
\cos\eta=\cos\phi_{i}\cos\phi_{s}+\cos\Delta\vartheta_{is}\sin\phi_{i}\sin\phi_{s},
\label{eqn:5}
\end{equation} 
where $\Delta\vartheta_{is} = \vartheta_{i} - \vartheta_{s}$ is the difference in phase-matching angles of the idler and signal photons.

The scattering function $S(\theta_{i},\theta_{s};E_{oi},E_{os},Z_{i},Z_{s})$, in Eqn. (\ref{eqn:3a}) and (\ref{eqn:3b}), takes into account deviations from the Klein-Nishina model when the idler and signal photons each scatter off an electron bound to atoms with atomic number $Z_{i}$ and $Z_{s}$, respectively. If the scattering medium of the idler and signal photons are space-like separated and are composed of identical elements with atomic number $Z$, then $S(\theta_{i},$ $ \theta_{s};$ $ E_{oi},$ $ E_{os},$ $ Z_{i},$ $ Z_{s})$ is assumed to be factorizable such that
\begin{equation}
	S(\theta_{i},\theta_{s};E_{oi},E_{os},Z,Z)=S_{i}(\theta_{i};E_{oi},Z)S_{s}(\theta_{s};E_{os},Z),
	\label{eqn:6}
\end{equation}
and Eq. (\ref{eqn:3a}) and (\ref{eqn:3b}) can be expressed as
\begin{subequations}
	\begin{equation}
		\begin{aligned}
			\frac{\partial^{2}\sigma_{incoh}^{\left(\Psi_{l}^{\pm}\right)}}{\partial\Omega_{i}\partial\Omega_{s}}=\frac{\partial^{2}\sigma_{KN}^{\left(\Psi_{l}^{\pm}\right)}}{\partial\Omega_{i}\partial\Omega_{s}}S_{i}(\theta_{i};E_{oi},Z)S_{s}(\theta_{s};E_{os},Z),
			\label{eqn:7a}
		\end{aligned}
	\end{equation}
	and	
	\begin{equation}
		\begin{aligned}
			\frac{\partial^{2}\sigma_{incoh}^{\left(\Phi_{l}^{\pm}\right)}}{\partial\Omega_{i}\partial\Omega_{s}}=\frac{\partial^{2}\sigma_{KN}^{\left(\Phi_{l}^{\pm}\right)}}{\partial\Omega_{i}\partial\Omega_{s}}S_{i}(\theta_{i};E_{oi},Z)S_{s}(\theta_{s};E_{os},Z).
		\end{aligned}
		\label{eqn:7b}
	\end{equation}
	\label{eqn:7}
\end{subequations}

In the analysis that follows, theoretical results are obtained using the work by Shwartz and Harris who found phase-matching angles in the case of degenerate 12.5 keV and non-degenerate 10 and 15 keV Bell pairs~\cite{Shwartz2011}. 

To make accurate theoretical Compton scattering predictions in the hard X-ray regime, it is important to select a scattering medium that satisfies the condition $E_{b} << E_{o}$. For photons in the $10^{-2}$ MeV range, we have selected neutral helium $(Z=2)$, since $E_{b}=0.0547$ keV and Compton scattering is the dominant reaction channel at this energy range when $Z < 5$ (See Fig. (18) in ref.~\cite{Bergstrom1997}). We use the values of $S$ for helium calculated by Brown~\cite{Brown1970}, since the total ground state energy of the wave functions used to derive $S$ only differ by $0.02\%$ from the most accurate ground state energy values. 

For incident energies above the K-shell ionization threshold, around 90 percent of scattering occurs from K-shell electrons~\cite{Bergstrom1997}. If both electrons of each scatterer is ejected from the K-shell, it is reported that the dependence of their angular distribution on the relative azimuth $\eta$ is proportional to $\sin^{2}\eta$, which is maximum for perpendicular azimuths and zero for coplanar azimuths. Therefore, it is expected that some 65 percent of coincidences will correspond to both electrons originating in the K-shell. One may therefore expect to also observe these azimuthal correlations in Compton scattering experiments~\cite{Pryce1947}.
\sectionn{Theoretical precision of angular distributions of hard X-rays}
\label{sec:Section4}
The scattering function $S$ is based on the impulse approximation (IA) model~\cite{Hubbell1975}. In this model, bound electrons are replaced by free electrons with a momentum distribution determined by the bound electron wave functions~\cite{Bergstrom1997}. The validity of the IA approximation can be understood in terms of the uncertainty relation in which it is difficult to distinguish between bound and free electrons due to the short time scale of a Compton interaction, which is no greater than $10^{-11}$ seconds~\cite{Bergstrom1997,Bay1955}. 

The electron binding energy $E_{b}$ of an electronic shell is included in the IA model only insofar that the binding energy places a kinematic limit on the lowest incident photon energy that can inelastically scatter from a bound electron~\cite{Bergstrom1997}. The binding energy  $E_{b}$ also places a limit on the highest scattered photon energy $E_{max}$ that can be observed for a given photon incident energy $E_{o}$. This limit is $E_{max}=E_{o} – E_{b}$~\cite{Bergstrom1997,Eisenberger1970}. Taking into account electron binding energy, the scattered photon energy $E$ is defined by the Compton formula
\begin{equation}
E = \frac{E_{max}}{1 + E_{max}(1 - \cos\theta)},
\label{eqn:8}
\end{equation}
where all energies are defined in units of $511$ keV (i.e., $mc^{2}$). 

For a single photon beam scattering off electrons from a many electron system, it has been shown that the IA model is accurate to order of $[E_{b}/(E_{o} - E)]^{2}$~\cite{Eisenberger1970,Williams1977}. The fractional precision of the theoretical incoherent cross section, defined here as $A(E_{oi}, $ $E_{os}, $ $E_{i}, $ $E_{s}, $ $E_{b})$, is derived using the rules of error propagation, such that the fractional precision $[E_{b}/(E_{oi}-E_{i})]^{2}$ associated with the incoherent scattering of the idler, and $[E_{b}/(E_{os}-E_{s})]^{2}$ associated with the signal, add in quadrature such that

\begin{equation}
	A\left(E_{oi}, E_{os}, E_{i}, E_{s}, E_{b}\right) = \sqrt{\left(\frac{E_{b}^{2}}{(E_{oi}-E_{i})^{2}}\right)^{2}+\left(\frac{E_{b}^{2}}{(E_{os}-E_{s})^{2}}\right)^{2}},
	\label{eqn:9}
\end{equation}
where the energies of the scattered photons $E_{i}$ and $E_{s}$ are defined using Eq. (\ref{eqn:8}) such that
\begin{equation}
	E_{i} = \frac{E_{oi}-E_{b}}{1 + (E_{oi}-E_{b})(1 - \cos\theta_{i})},\quad E_{s} = \frac{E_{os}-E_{b}}{1 + (E_{os}-E_{b})(1 - \cos\theta_{s})}.
	\label{eqn:10}
\end{equation}
Therefore, the incoherent cross sections for $\ket{\Psi_{l}^{\pm}}$ and $\ket{\Phi_{l}^{\pm}}$ which takes into account the theoretical precision are expressed as follows 
\begin{subequations}
	\begin{equation}
		\begin{aligned}
			\frac{\partial^{2}\sigma_{incoh}^{\left(\Psi_{l}^{\pm}\right)}}{\partial\Omega_{i}\partial\Omega_{s}}=\frac{\partial^{2}\sigma_{KN}^{\left(\Psi_{l}^{\pm}\right)}}{\partial\Omega_{i}\partial\Omega_{s}}S_{i}(\theta_{i};E_{oi},Z)S_{s}(\theta_{s};E_{os},Z)\pm\frac{1}{2}\Delta\sigma^{\left(\Psi_{l}^{\pm}\right)}_{(IA)}(\theta_{i},\theta_{s};E_{oi},E_{os},Z),
			\label{eqn:11a}
		\end{aligned}
	\end{equation}
	and	
	\begin{equation}
		\begin{aligned}
			\frac{\partial^{2}\sigma_{incoh}^{\left(\Phi_{l}^{\pm}\right)}}{\partial\Omega_{i}\partial\Omega_{s}}=\frac{\partial^{2}\sigma_{KN}^{\left(\Phi_{l}^{\pm}\right)}}{\partial\Omega_{i}\partial\Omega_{s}}S_{i}(\theta_{i};E_{oi},Z)S_{s}(\theta_{s};E_{os},Z)\pm\frac{1}{2}\Delta\sigma^{\left(\Phi_{l}^{\pm}\right)}_{(IA)}(\theta_{i},\theta_{s};E_{oi},E_{os},Z),
		\end{aligned}
		\label{eqn:11b}
	\end{equation}
	\label{eqn:11}
\end{subequations}
where the theoretical uncertainty of the cross section $\Delta\sigma_{(IA)}$ for the states $\ket{\Psi_{l}^{\pm}}$ and $\ket{\Phi_{l}^{\pm}}$ are defined, respectively, as 
\begin{subequations}
	\begin{equation}
		\Delta\sigma^{\left(\Psi_{l}^{\pm}\right)}_{(IA)}(\theta_{i},\theta_{s};E_{oi},E_{os},Z)=\frac{\partial^{2}\sigma_{KN}^{\left(\Psi_{l}^{\pm}\right)}}{\partial\Omega_{i}\partial\Omega_{s}}S_{i}(\theta_{i};E_{oi},Z)S_{s}(\theta_{s};E_{os},Z)A\left(E_{oi}, E_{os}, E_{i}, E_{s}, E_{b}\right),
		\label{eqn:12a}
	\end{equation}
	and
	\begin{equation}
		\Delta\sigma^{\left(\Phi_{l}^{\pm}\right)}_{(IA)}(\theta_{i},\theta_{s};E_{oi},E_{os},Z)=\frac{\partial^{2}\sigma_{KN}^{\left(\Phi_{l}^{\pm}\right)}}{\partial\Omega_{i}\partial\Omega_{s}}S_{i}(\theta_{i};E_{oi},Z)S_{s}(\theta_{s};E_{os},Z)A\left(E_{oi}, E_{os}, E_{i}, E_{s}, E_{b}\right).
		\label{eqn:12b}
	\end{equation}
	\label{eqn:12}
\end{subequations}
The theoretical precision is improved as $\Delta\sigma_{(IA)}\mapsto0$, which can happen in one of two ways: If one is constrained to use a particular scattering medium, then the incident photon energy is chosen such that $E_{o} >> E_{b}$. Or, if one is constrained to a given photon energy $E_{o}$, then a scattering medium with sufficiently low binding energy is chosen such that $E_{b} $ $<< $ $E_{o}$.
\sectionn{The counting rates $N_{\perp}$ and $N_{\parallel}$}
\label{sec:Section5}
Compton scattering of polarization entangled photons has been studied in the past by measuring a quantity known as the azimuthal ratio. This involves a coincidence measurement of both photons scattered through the same angle, i.e., $\theta_{s}=\theta_{i}=\theta$~\cite{Pryce1947,Snyder1948}. The geometrical arrangement to measure the ratio of cross-polarized Bell states when the wave vectors are not colinear is illustrated schematically in Fig. (\ref{fig:Fig2}). In this arrangement, the photon counter of the scattered idler is fixed at $\phi_{i}=0^{o}$, such that only idler photons scattered in the trajectory plane are detected. Using Eqn. (\ref{eqn:5}), this implies $\eta = \phi_{s}$.

The quantity $N_{\perp}$ is defined as the coincidence count rate when the photon counter of the signal is set at $\phi_{s}=90^{o}$, such that $\eta=\eta_{(max)}=90^{o}$. The quantity $N_{\parallel}$ is the coincidence count rate when the signal photon counter is set to $\phi_{s}=0^{o}$, and therefore $\eta=\eta_{(min)}=0^{o}$, and the ratio of these quantities is then determined. The count rates $N_{\perp}$ and $N_{\parallel}$ are proportional to the incoherent cross sections, which can be expressed as follows
\begin{subequations}
	\begin{equation}
	N_{\perp} \propto \frac{\partial^{2}\sigma_{incoh}}{\partial\Omega_{i}\partial\Omega_{s}}\left(\theta,\eta=90^{o};E_{oi},E_{os},Z\right),
	\label{eqn:13a}
	\end{equation}
and
	\begin{equation}
	N_{\parallel} \propto \frac{\partial^{2}\sigma_{incoh}}{\partial\Omega_{i}\partial\Omega_{s}}\left(\theta,\eta=0^{o};E_{oi},E_{os},Z\right).
	\label{eqn:13b}
\end{equation}
\label{eqn:13}
\end{subequations}
Before evaluating the ratios, which is done in the next two sections, it is
instructive to plot Eqn. (\ref{eqn:13a}) and (\ref{eqn:13b}) and examine both their structure and the theoretical precision as a function of
$\theta$. These cross sections, normalized over the classical electron 
\begin{framed}
	\begin{center}
		\includegraphics[width=\linewidth]{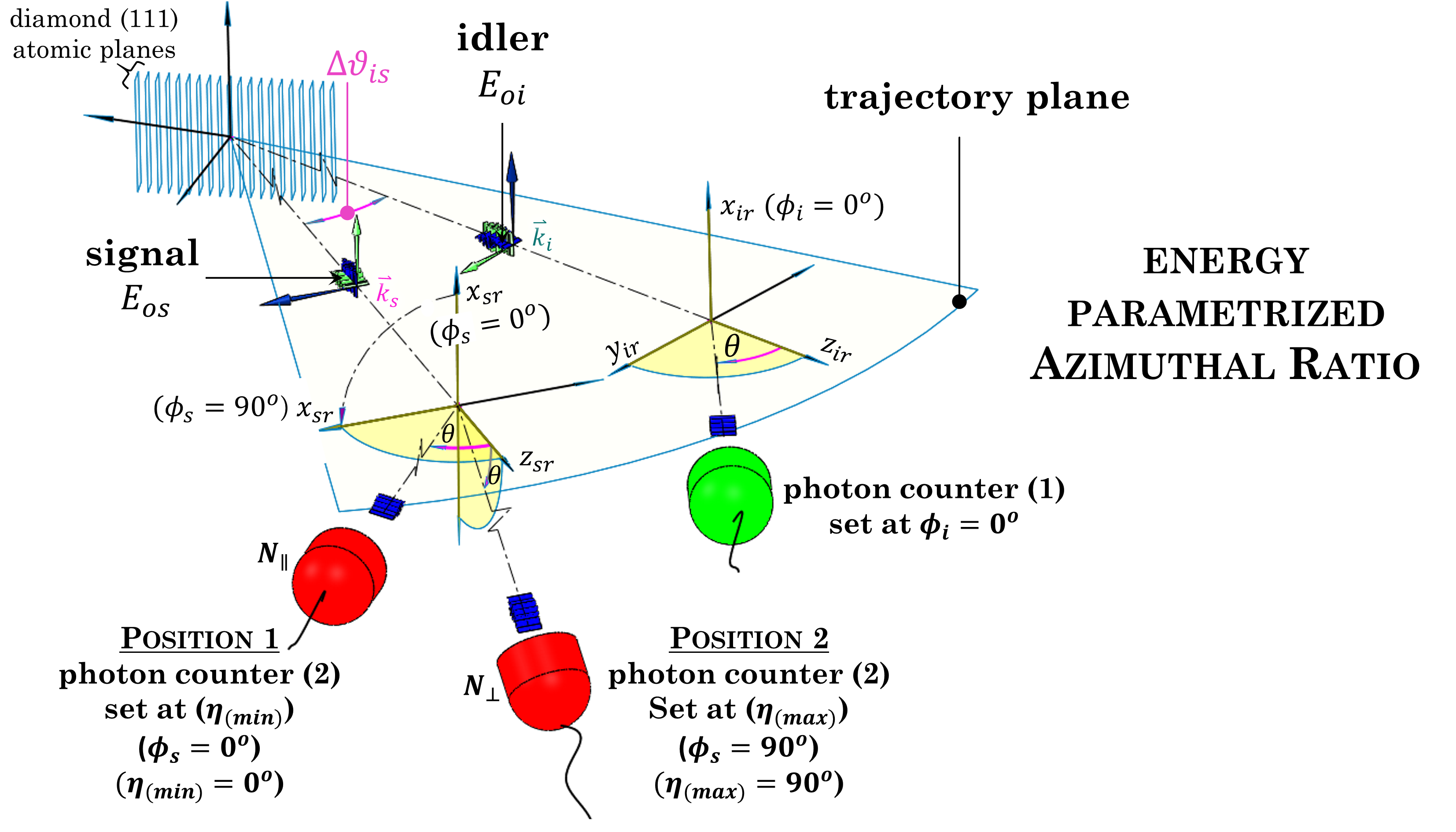}
		\captionof{figure}{The geometrical arrangement of an energy parameterized azimuthal ratio measurement. The measurement assumes the counter of the idler photon remains fixed at an azimuthal angle of $\phi_{i}=0^{o}$ and the counter of the signal photon is placed at values of $\eta_{(min)}=0^{o}$ ($\phi_{s}= 0^{o}$) and $\eta_{(max)}=90^{o}$ ($\phi_{s}= 90^{o}$).} 
		\label{fig:Fig2}
	\end{center}
\end{framed}
radius ($r_{0}^{4}$), have been evaluated and plotted in Fig.
(\ref{fig:Fig3}a) for degenerate 12.5 keV photons in the $\ket{\Psi^{\pm}_{l}}$ states. As can be seen, there is a significant
deviation between the cross sections in the angular range $80^{o} \le \theta \le 100^{o}$, where the cross section evaluated at $\eta = 0^{o}$ drops by around three orders of magnitude relative to the cross section evaluated at $\eta = 90^{o}$. The shaded area gives the estimated theoretical precision. Within the angular range $80^{o} \le \theta \le 100^{o}$, the estimated theoretical precision is around 8$\%$ for both cross sections.   

For the particular case that $\eta_{(min)} =0^{o}$ and $\eta_{(max)} = 90^{o}$, an equality is found in the cross sections for Compton scattering of cross- and parallel-polarization entangled Bell states. Using Eqns. given in (\ref{eqn:4}) and (\ref{eqn:11}) it can be shown that
\begin{subequations}
	\begin{equation}
	 \frac{\partial^{2}\sigma_{incoh}^{\left(\Psi^{\pm}_{l}\right)}}{\partial\Omega_{i}\partial\Omega_{s}}\left(\theta,\eta_{(max)};E_{oi},E_{os},Z\right)=\frac{\partial^{2}\sigma_{incoh}^{\left(\Phi^{\pm}_{l}\right)}}{\partial\Omega_{i}\partial\Omega_{s}}\left(\theta,\eta_{(min)};E_{oi},E_{os},Z\right),
	 \label{eqn:14a}
	\end{equation}
and
	\begin{equation}
	\frac{\partial^{2}\sigma_{incoh}^{\left(\Psi^{\pm}_{l}\right)}}{\partial\Omega_{i}\partial\Omega_{s}}\left(\theta,\eta_{(min)};E_{oi},E_{os},Z\right)=\frac{\partial^{2}\sigma_{incoh}^{\left(\Phi^{\pm}_{l}\right)}}{\partial\Omega_{i}\partial\Omega_{s}}\left(\theta,\eta_{(max)};E_{oi},E_{os},Z\right).
	\label{eqn:14b}
\end{equation}
\label{eqn:14}
\end{subequations}

Figure (\ref{fig:Fig3}b) gives a plot of the cross sections for nondegenerate 10 and 15 keV cross- and parallel-polarization entangled states. 

\sectionn{Energy parameterized azimuthal ratios}
\label{sec:Section6}
\begin{framed}
	\begin{center}
		\includegraphics[width=\linewidth]{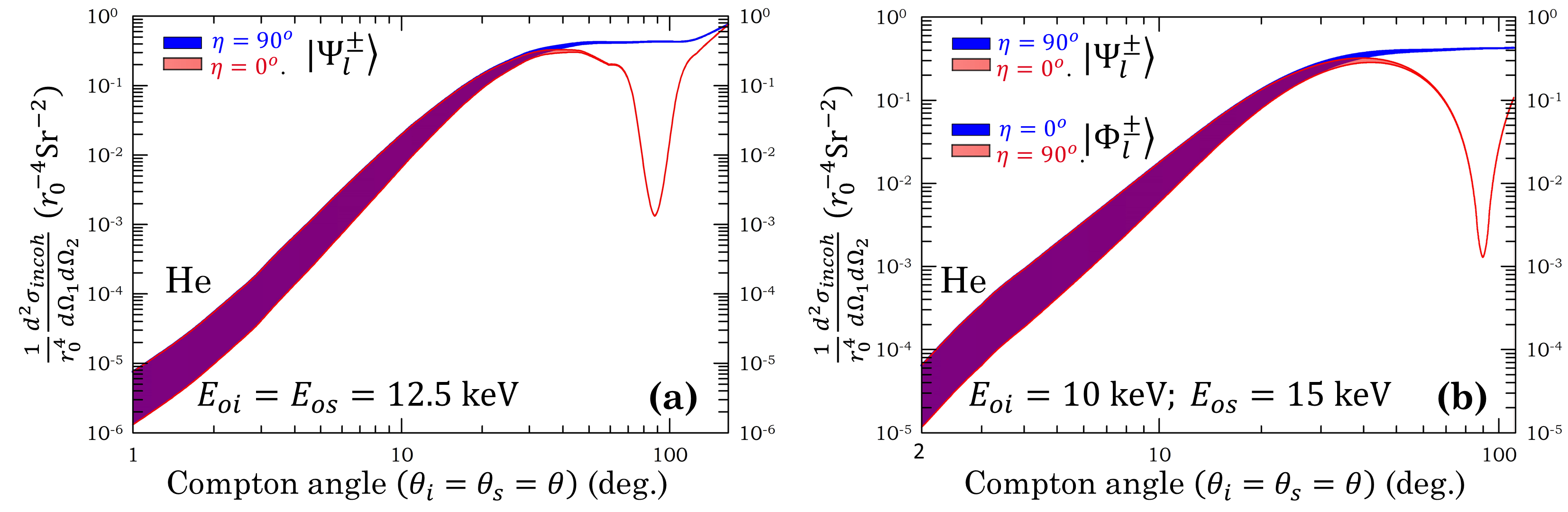}
		\captionof{figure}{Log-log plot of the incoherent differential cross sections for degenerate (a) and nondegenerate (b) states evaluated at $\eta = 0^{o}$ ($\phi_{s}=0^{o}$) and $\eta = 90^{o}$ ($\phi_{s}=90^{o}$), when $\phi_{i}=0^{o}$. The shaded area is a first order estimate of the uncertainty in the theoretical cross section when helium is used as a scattering medium. For this case, the uncertainty in the cross section is around $8\%$ between $80^{o}\le\theta\le 100^{o}$. (Photon energies and phase-matching angles are obtained from reference~\cite{Shwartz2011}.)} 
		\label{fig:Fig3}
	\end{center}
\end{framed}
The angle $\Delta\vartheta_{is} = \vartheta_{i} - \vartheta_{s}$ is the difference in phase-matching angles, which were introduced in Sec. (\ref{sec:Section2}). As shown in Fig. (\ref{fig:Fig2}), $\Delta\vartheta_{is}$ is the angle subtended between $\vectarrow{k}_{s}$ and
$\vectarrow{k}_{i}$. An energy {\it{only}} parameterized azimuthal ratio is defined for coincidence measurements in which $\eta$ is independent of $\Delta\vartheta_{is}$. The setup shown in Fig. (\ref{fig:Fig2}) satisfies this definition. Therefore, the expected count rates $N_{\perp}$ and $N_{\parallel}$ as a function of $\theta$ will be invariant to changes in $\Delta\vartheta_{is}$, which implies that the ratio of these quantities will be invariant as well.  

The expected energy {\it{only}} dependent ratio for cross-polarization entangled states that takes into account the theoretical precision is defined as follows 
\begin{equation}
\frac{N^{\left(\Psi^{\pm}_{l}\right)}_{\perp}}{N^{\left(\Psi^{\pm}_{l}\right)}_{\parallel}} = R_{nd}^{\left(\Psi^{\pm}_{l}\right)}\left(\theta;E_{oi},E_{os}\right)\pm\frac{\sqrt{2}}{2}R_{nd}^{\left(\Psi^{\pm}_{l}\right)}\left(\theta;E_{oi},E_{os}\right)A\left(E_{oi}, E_{os}, E_{i}, E_{s}, E_{b}\right),
\label{eqn:15}
\end{equation}
where we have applied error propagation methods to determine the theoretical precision. The subscript $nd$ in $R$ denotes `nondegenerate' energies. Using Eqn. (\ref{eqn:3a}) and Eqn. (\ref{eqn:4a}) evaluated at $\eta_{(max)} = 90^{o}$ and $\eta_{(min)} = 0^{o}$, $R_{nd}$ is given by 
\begin{equation}
\begin{aligned}
	R_{nd}^{\left(\Psi^{\pm}_{l}\right)}\left(\theta;E_{oi},E_{os}\right)&=\frac{\partial^{2}\sigma_{incoh}^{\left(\Psi^{\pm}_{l}\right)}\left(\theta,\eta_{(max)};E_{oi},E_{os}\right)}{\partial^{2}\sigma_{incoh}^{\left(\Psi^{\pm}_{l}\right)}\left(\theta,\eta_{(min)};E_{oi},E_{os}\right)}\\
	&=\frac{\sum\limits_{a=0}^{1}\bra{I}T\left(\theta_{i};E_{oi}\right)M\left(\frac{a\pi}{2}\right)\ket{+}\bra{I}T\left(\theta_{s};E_{os}\right)M\left(\frac{a\pi}{2}\right)\ket{+}}{\sum\limits_{a=0}^{1}\bra{I}T\left(\theta_{i};E_{oi}\right)M\left(\frac{a\pi}{2}\right)\ket{+}\bra{I}T\left(\theta_{s};E_{os}\right)M\left(\frac{a\pi}{2}\right)\ket{-}},
\end{aligned}
\label{eqn:16}
\end{equation} 
where the scattering function $S$ cancels due to the definition of the incoherent cross section given in Eqn. (\ref{eqn:3}).

In the case for degenerate energies, i.e., $E_{os}=E_{oi}=E_{o}$, Eqn. (\ref{eqn:16}) simplifies to 
\begin{equation}
	R_{d}^{\left(\Psi^{\pm}_{l}\right)}\left(\theta;E_{o}\right)=\frac{\bra{I}T\left(\theta;E_{o}\right)\ket{+}^{2}+\bra{I}T\left(\theta;E_{o}\right)\ket{-}^{2}}{2\bra{I}T\left(\theta;E_{o}\right)\ket{+}\bra{I}T\left(\theta;E_{o}\right)\ket{-}},
	\label{eqn:17}
\end{equation}
\begin{framed}
	\begin{center}
		\includegraphics[width=\linewidth]{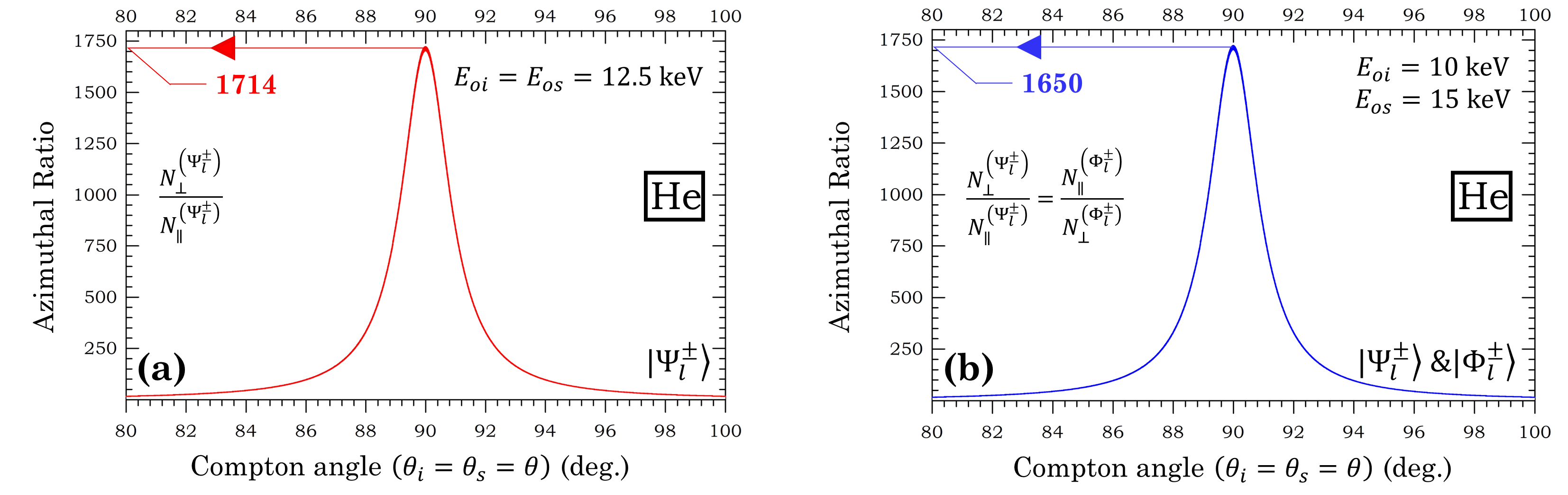}
		\captionof{figure}{Energy parameterized azimuthal ratios. \textbf{(a)}: The expected azimuthal ratio in the case of $\ket{\Psi^{\pm}_{l}}$ states having degenerate incident energies of $E_{oi}=E_{os}=E_{o}=(12.5~\mbox{keV})/(511~\mbox{keV})=0.0245$. \textbf{(b)}: The azimuthal ratio of $\ket{\Psi^{\pm}_{l}}$ and $\ket{\Phi^{\pm}_{l}}$ in the case of non-degenerate incident energies $E_{os}=1.2E_{o}$ and $E_{oi}=0.8E_{o}$. (Photon energies and phase-matching angles are obtained from reference~\cite{Shwartz2011}.)} 
		\label{fig:Fig4}
	\end{center}
\end{framed}
where the subscript `$d$' denotes `degenerate' energy. 

Using equations of (\ref{eqn:14}), the energy dependent ratios for parallel-polarization entangled Bell states are related to the ratios for cross polarization entangled states as follows 
\begin{equation}
\begin{aligned}
R_{nd}^{\left(\Phi^{\pm}_{l}\right)}\left(\theta;E_{oi},E_{os}\right) = \frac{1}{R_{nd}^{\left(\Psi^{\pm}_{l}\right)}\left(\theta;E_{oi},E_{os}\right)}.
\end{aligned}
	\label{eqn:18}
\end{equation}

The ratio $R_{d}^{\left(\Psi^{\pm}_{l}\right)}\left(\theta; E_{o}\right)$ for $E_{o} = (12.5~\mbox{keV})/(511~\mbox{keV}) =0.0245$ has been calculated and is shown in Fig. \ref{fig:Fig4}(a). The thickness of the curve gives an estimate of the theoretical precision. The ratios $R_{nd}^{\left(\Psi^{\pm}_{l}\right)}$ and $R_{nd}^{\left(\Phi^{\pm}_{l}\right)}$ have also been calculated for the case when $E_{os} = 1.2E_{o}$ (i.e., $E_{os} = 15$ keV) and  $E_{oi} = 0.8E_{o}$ (i.e., $E_{oi} = 10$ keV), and are shown in Fig. \ref{fig:Fig4}(b). As can be seen, both Fig. \ref{fig:Fig4}(a) and \ref{fig:Fig4}(b) show a significant asymmetry in the coincidence counting rates at $\theta\approx 89.98$ degrees, where the degenerate and non-degenerate ratios reach a peak value of $1713.04$ and $1650.26$, respectively. These peak values are approximately $600$ times larger compared to the value of 2.85 predicted for 511 keV cross-polarized Bell states~\cite{Pryce1947,Snyder1948}.

\sectionn{Azimuthal ratios parameterized by the incident energies and phase-matching angles}
\label{sec:Section7}
{ \fontfamily{times}\selectfont
\noindent
We now propose a coincidence measurement of the azimuthal ratio parameterized in terms of incident energies $E_{os}$ and $E_{oi}$, and the difference in phase matching angles $\Delta\vartheta_{is}=\vartheta_{i}-\vartheta_{s}$. (The phase matching angles used here are obtained from~\cite{Shwartz2011}). The geometrical arrangement is illustrated schematically in Fig. (\ref{fig:Fig5}). As can be seen, the photon counter measuring the scattered signal photons is set at the same two positions as in Fig. (\ref{fig:Fig2}).

However, in this scheme, photon counter (1) is now fixed at right angles to the trajectory plane (i.e., $\phi_{i}=90^{o}$), rather than parallel to it, which was the case for energy parameterized ratios discussed in Sec. (\ref{sec:Section6}). Using Eqn. (\ref{eqn:5}), we find that $\cos\eta = \cos\Delta\vartheta_{is}\sin\phi_{s}$. When photon counter (2) is at position (1), ($\phi_{s}=0^{o}$), the two scattering planes of the signal is perpendicular to that of the idler, and therefore, ($\eta=\eta_{(max)}=90^{o}$). When the photon counter (2) is set at position (2), ($\phi_{s}=90^{o}$), it can be shown that $\eta_{(min)}=\Delta\vartheta_{is}$.
\begin{framed}
	\begin{center}
		\includegraphics[width=\linewidth]{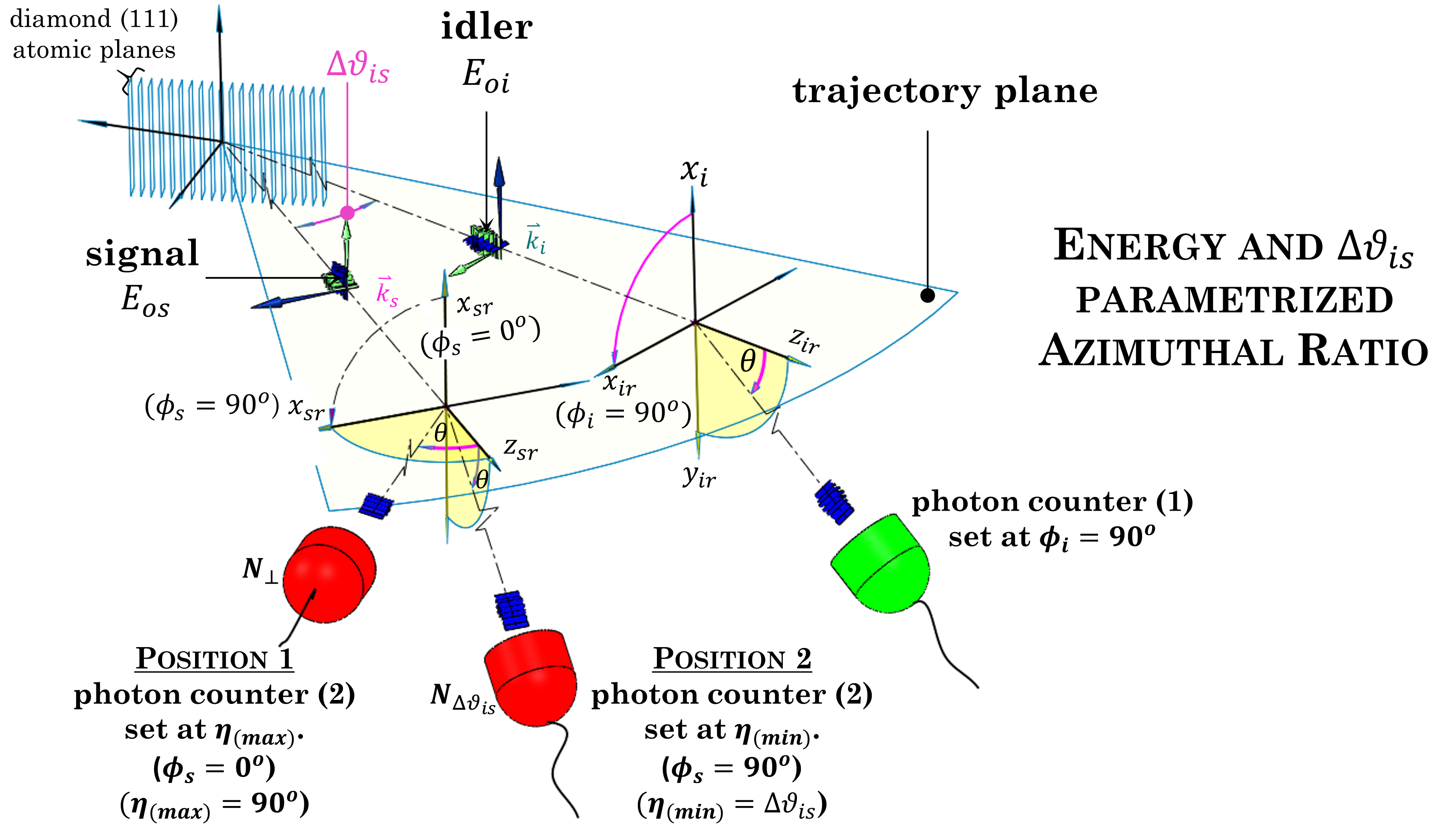}
		\captionof{figure}{The geometrical arrangement of an energy and $\Delta\vartheta_{is}$ parameterized azimuthal ratio measurement. The measurement assumes the counter of the idler photon remains fixed at an azimuthal angle $\phi_{i}=90^{o}$, and the counter of the signal photon is placed at values of $\eta_{(min)}=\Delta\vartheta_{is}$ ($\phi_{s}=90^{o}$) and $\eta_{(max)}=90^{o}$ ($\phi_{s}=0^{o}$). {\textit{Refer to Eqn. (\ref{eqn:5}).}}} 
		\label{fig:Fig5}
	\end{center}
\end{framed}

We now define $\rho_{nd}(\theta;E_{oi}, E_{os},\Delta\vartheta_{is})$ as the energy and phase-matching parameterized azimuthal ratio for the state $\ket{\Psi^{\pm}_{l}}$, such that
\begin{equation}
\frac{N^{\left(\Psi^{\pm}_{l}\right)}_{\perp}}{N^{\left(\Psi^{\pm}_{l}\right)}_{\Delta\vartheta_{is}}} =\rho_{nd}^{\left(\Psi^{\pm}_{l}\right)}\left(\theta;E_{oi},E_{os},\Delta\vartheta_{is}\right)\pm\frac{\sqrt{2}}{2}\rho_{nd}^{\left(\Psi^{\pm}_{l}\right)}\left(\theta;E_{oi},E_{os},\Delta\vartheta_{is}\right)A\left(E_{oi}, E_{os}, E_{i}, E_{s}, E_{b}\right),
\label{eqn:19}
\end{equation}
where
\begin{equation}
\rho_{nd}^{\left(\Psi^{\pm}_{l}\right)}\left(\theta;E_{oi},E_{os},\Delta\vartheta_{is}\right)=\frac{\sum\limits_{a=0}^{1}\bra{I}T\left(\theta_{i};E_{oi}\right)M\left(\frac{a\pi}{2}\right)\ket{+}\bra{I}T\left(\theta_{s};E_{os}\right)M\left(\frac{a\pi}{2}\right)\ket{+}}{\sum\limits_{a=0}^{1}\bra{I}T\left(\theta_{i};E_{oi}\right)M\left(\frac{a\pi}{2}\right)\ket{+}\bra{I}T\left(\theta_{s};E_{os}\right)M\left(\Delta\vartheta_{is}+\frac{a\pi}{2}\right)\ket{-}},
\label{eqn:20}
\end{equation}
and applying the equalities found in Eqns. (\ref{eqn:14}), we can define the ratio for the state $\ket{\Phi^{\pm}_{l}}$ as 
\begin{equation}
 \rho_{nd}^{\left(\Phi^{\pm}_{l}\right)} = \frac{1}{\rho_{nd}^{\left(\Psi^{\pm}_{l}\right)}}.
 \label{eqn:22}
\end{equation}

Equations (\ref{eqn:19}) has been calculated and plotted in Fig. (\ref{fig:Fig6}), when $\Delta\vartheta_{is} = 78.38^{o}, 81.53^{o}$ for the $\ket{\Phi^{\pm}_{l}}$ states with non-degenerate energies $E_{os}= 1.2E_{o}$ and $E_{oi}= 0.8E_{o}$.

\sectionn{Conclusion}
{ \fontfamily{times}\selectfont
\noindent
Technology can now produce all 4 Bell states with sufficient incident energies required for Compton scattering experiments. Relative to 511 keV cross-polarized photons,
\begin{framed}
	\begin{center}
		\includegraphics[width=\linewidth]{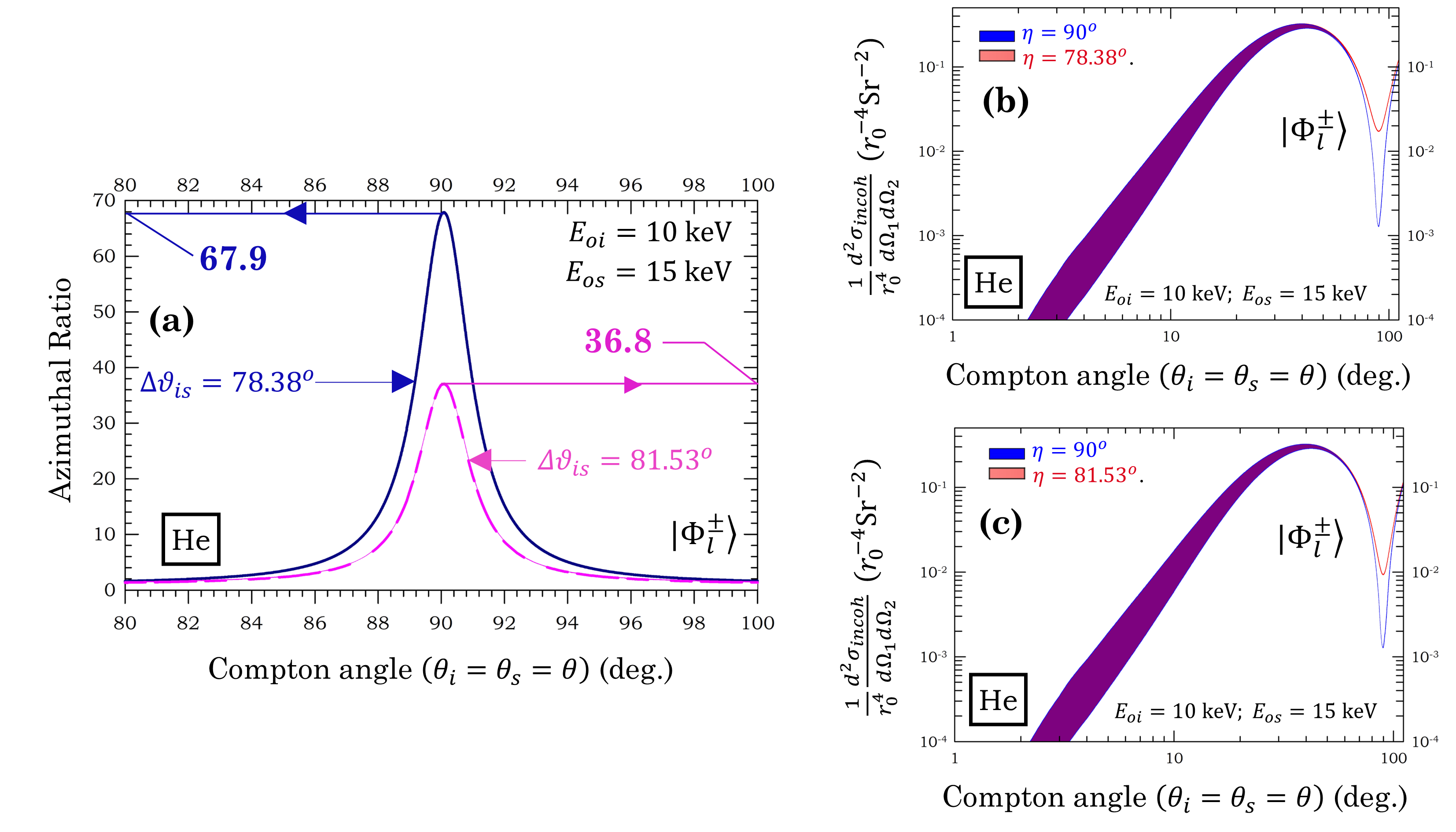}
		\captionof{figure}{\textbf{(a)}: Energy and $\Delta\vartheta_{is}$ parameterized azimuthal ratios. The ratio for parallel-polarization entangled Bell states in the case of nondegenerate incident energies $E_{os}=1.2E_{o}$ and $E_{oi}=0.8E_{o}$, with phase-matching angle $\Delta\vartheta_{is}=78.38^{o}$ and $\Delta\vartheta_{is}=81.53^{o}$. \textbf{(b)} and \textbf{(c)}: Log-log plot of the inchorent differential cross sections used to calcuate the ratios in Fig. (a). (Photon energies and phase-matching angles are obtained from reference~\cite{Shwartz2011}.)} 
		\label{fig:Fig6}
	\end{center}
\end{framed}
hard X-ray Bell states with degenerate or non-degenerate incident energies show a significant amplification in polarization correlations, making them amenable to experimental verification, provided a suitable scattering medium is selected. 

When all 4 Bell states are considered, relationships in the cross sections and ratios are found between cross and parallel-polarization entangled states. In addition, the azimuthal ratio can be further parameterized in terms of the phase matching angles of Bell state photons generated in parametric down-conversion processes. In principle, these mathematical relationships and phase-matching constraints each provide a stringent test of
Compton scattering theory itself. Therefore, the present study may be useful for future testing of fundamental laws of quantum mechanics.

Conducting a systematic investigation of this type could validate conclusions of past fundamental and practical research activities. Furthermore, if future experiments confirm theory, then Compton scattering could be used, aside from polarization filters, as a tool to verify the production of high energy polarization-entangled states emitted at specific phase-matching angles.

In future works, we will investigate more practical applications of these polarization-entangled states in the form of signal communication and extracting information in the field of medical imaging. On a fundamental
level, since the matrix formalism presented here can also incorporate the dynamics of polarized Compton electrons, a study will be conducted on how entanglement is shared between the Bell state photons and the Compton electrons.     
  
}

{\color{maroon}
\vskip 6mm
 \noindent\Large\bf Acknowledgments}
 \vskip 3mm
{ \fontfamily{times}\selectfont
 \noindent
I would like to thank Tadayuki Takahashi for his helpful
discussions. This research was supported by JSPS KAKENHI Grant Number JP19K23436, the Grant-in-Aid for Scientific Research on Innovative Areas ``Toward new frontiers: Encounter and synergy of state-of-the-art astronomical detectors and exotic quantum beams'', JSPS KAKENHI grant numbers JP16H02170, JP18H05457 and JP18H02700, and  the Grant-in-Aid for Scientific Research (A), Grant Number JP20H00153.}

{\color{maroon}


\begin{thebibliography}{10}
	{\color{black}
		{
		\bibitem{McNamara2014}
		McNamara A L, Toghyani M, Gillam J E, Wu K, and Kuncic Z 2014
		\newblock {\em Towards optimal imaging with PET: an in silico feasibility study}
		\newblock {\em \it{Physics in medicine and biology}}, \href{https://www.ncbi.nlm.nih.gov/pubmed/25415271}{{\bf{59}}(24) 7587--7600}
		\bibitem{Kasday1971}
		Kasday L R 1971
		\newblock {\em Rendiconti S.I.F, Corso IL}
		\newblock {\em {\it{New York, Academic Press, 1971}}, \href{}{{\bf{}} (pg. 198--210)}}	
		\bibitem{Kasday1975}
		Kasday L R, Ullman J D and Wu C S 1971-1996
		\newblock {\em Angular correlation of compton-scattered annihilation photons and hidden variables}
		\newblock {\em {\it{Il Nuovo Cimento B}}}, \href{https://doi.org/10.1007/BF02724742}{{\bf{25}}(2) 633--661}
		\bibitem{Einstein1935}
		Einstein A, Podolsky Band Rosen N 1935
		\newblock {\em Can Quantum-Mechanical Description of Physical Reality Be Considered Complete?}
		\newblock {\em {\it{American Physical Society}}}, 
		\href{https://link.aps.org/doi/10.1103/PhysRev.47.777}{{\bf{47}} (777--780)}
		\bibitem{Bohm1957}
		Bohm D and Aharonov Y 1957
		\newblock {\em Discussion of Experimental Proof for the Paradox of Einstein, Rosen, and Podolsky}
		\newblock {\em {\it{Phys.Rev.}}}, \href{https://link.aps.org/doi/10.1103/PhysRev.108.1070}{{\bf{108}} (1070--1076)}
		\bibitem{Bohm1976}
		Bohm D J and Hiley B J 1976
		\newblock {\em Nonlocality and polarization correlations of annihilation quanta}
		\newblock {\em {\it{Il Nuovo Cimento B (1971-1996)}}}, \href{https://doi.org/10.1007/BF02726290}{{\bf{35}} 137--144}
		\bibitem{Heitler1954}
		Heitler W 1954
		\newblock {\em The quantum theory of radiation}
		\newblock {\em {\it{Oxford University Press}}, \href{https://store.doverpublications.com/0486645584.html}{{\bf{Third edition}} (223)}}
		\bibitem{Caradonna2019}
		Caradonna P and Reutens D and Takahashi T and Takeda S and Vegh V 2019
		\newblock {\em Probing entanglement in Compton interactions}
		\newblock {\em {\it{Journal of Physics Communications}}, \href{https://doi.org/10.1088\%2F2399-6528\%2Fab45db}{{\bf{3}} (105005)}}
		\bibitem{Hiesmayr2019}
		Hiesmayr B C and Moskal P 2019
		\newblock {\em Witnessing Entanglement In Compton Scattering Processes Via Mutually Unbiased Bases}
		\newblock {\em {\it{Scientific Reports}}}, \href{https://doi.org/10.1038/s41598-019-44570-z}{{\bf{9}} (8166)}
		\bibitem{Langhoff1960}
		Langhoff H 1960
		\newblock {\em Die Linearpolarisation der Vernichtungsstrahlung von Positronen}
		\newblock {\em {\it{Zeitschrift f{\"u}r Physik}}, \href{https://doi.org/10.1007/BF01336980}{{\bf{160}} (186--193)}}
		\bibitem{Bernhard2013}
		Bernhard W A and Christian B and Stefano M C and J{\"o}rg E and  Zolt{\'a}n H and Christoph H K and Adriana P and Antonio P and Ralf R and Yuri R and Kenji T 2013
		\newblock {\em X-ray quantum optics}
		\newblock {\em {\it{Journal of Modern Optics}},
		\href{https://doi.org/10.1080/09500340.2012.752113}{{\bf{60}}(1) 2--21}}
		\bibitem{Yabashi2001}
		Yabashi M and Mochizuki T and Yamazaki H and Goto S and Ohashi H and Takeshita K and Ohata T and Matsushita T and Tamasaku K and Tanaka Y and Ishikawa T 2001
		\newblock {\em Design of a beamline for the SPring-8 long undulator source 1}
		\newblock {\em {\it{Nuc. Instr. and Meth. in Phys. Res. Section A}}}, \href{http://www.sciencedirect.com/science/article/pii/S0168900201004442}{{\bf{467--468}} (678--681)}
		\bibitem{Shwartz2011}
		Shwartz S and Harris S E 2011
		\newblock {\em Polarization Entangled Photons at X-Ray Energies}
		\newblock {\em {\it{Phys. Rev. Lett.}}}, \href{https://link.aps.org/doi/10.1103/PhysRevLett.106.080501}{{\bf{106}} (080501)}				
		\bibitem{Ribberfors1982}
		Ribberfors R and Berggren K F 1982
		\newblock {\em Incoherent-x-ray-scattering functions and cross sections ${(\frac{d\ensuremath{\sigma}}{d{\ensuremath{\Omega}}^{\ensuremath{'}}})}_{\mathrm{incoh}}$ by means of a pocket calculator}
		\newblock {\em {\it{Phys. Rev. A}}}, \href{https://link.aps.org/doi/10.1103/PhysRevA.26.3325}{{\bf{26}} 3325--3333}	
		\bibitem{Wightman1948}
		Wightman A 1948
		\newblock {Note on Polarization Effects in Compton Scattering}
		\newblock {\em {\it{Phys. Rev.}}}, \href{https://link.aps.org/doi/10.1103/PhysRev.74.1813}{{\bf{74}}(12) 1813--1817}
		\bibitem{Fano1949}
		Fano U 1949
		\newblock {\em Remarks on the classical and quantum-mechanical treatment of partial polarization}
		\newblock {\em {\it{Journal of the Optical Society of America}}}, \href{https://www.osapublishing.org/josa/abstract.cfm?uri=josa-39-10-859}{{\bf{39}} 859--863}	
		\bibitem{McMaster1961}
		McMaster W H 1961
		\newblock {\em Matrix Representation of Polarization}
		\newblock {\em {\it{Reviews of Modern Physics}}}, \href{http://dx.doi.org/10.1103/RevModPhys.33.8}{{\bf{33}} 8}
		\bibitem{Bergstrom1997}
		Bergstrom P M and Pratt R H 1997
		\newblock {\em An overview of the theories used in compton scattering calculations}
		\newblock {\em {\it{Radiation Physics and Chemistry}}}, \href{https://doi.org/10.1016/S0969-806X(97)00022-4}{{\bf{50}} (1) 3--29}
		\bibitem{Brown1970}
		Brown Robert T 1970
		\newblock {\em Coherent and incoherent X-ray scattering by bound electrons. I. Helium isoelectronic sequence}
		\newblock {\em {\it{Physical Review A}}}, \href{https://link.aps.org/doi/10.1103/PhysRevA.1.1342}{{\bf{97}} (5) 1342--1347}	
		\bibitem{Pryce1947}
		Pryce M H L  and Ward J C 1947
		\newblock {\em Angular Correlation Effects with Annihilation Radiation}
		\newblock {\em {\it{Nature}}}, \href{https://doi.org/10.1038/160435a0}{{\bf{160}} 435}
		\bibitem{Hubbell1975}
		Hubbell J H  and Veigele Wm J and Briggs E A and Brown R T and Cromer D T and Howerton R J 1975
		\newblock {\em Atomic form factors, incoherent scattering functions, and photon scattering cross sections}
		\newblock {\em {\it{Journal of Physical and Chemical Reference Data}}}, \href{ https://doi.org/10.1063/1.555523}{{\bf{4}} (3) 471--538}
		\bibitem{Bay1955}
		Bay Z and Henri V P and McLernon F 1955
		\newblock {\em Simultaneity in the Compton Effect}
		\newblock {\em {\it{Phys. Rev.}}}, \href{https://link.aps.org/doi/10.1103/PhysRev.97.1710}{{\bf{97}} (6) 1710--1712}
		\bibitem{Eisenberger1970}
		Eisenberger P and Platzman P M 1970
		\newblock {\em Compton Scattering of X Rays from Bound Electrons}
		\newblock {\em {\it{Phys. Rev. A}}}, \href{https://link.aps.org/doi/10.1103/PhysRevA.2.415}{{\bf{2}} (2) 415--423}	
		\bibitem{Williams1977}
		Williams B 1977
		\newblock {\em Compton scattering}
		\newblock {\em {\it{McGraw-Hill, Inc.}}, \href{}{{\bf{}} (pg. 37)}}	
		\bibitem{Snyder1948}
		Snyder H S, Pasternack S and Hornbostel J 1948
		\newblock {Angular Correlation of Scattered Annihilation Radiation}
		\newblock {\em {\it{Phys. Rev.}}}, \href{https://journals.aps.org/pr/abstract/10.1103/PhysRev.73.440}{{\bf{73}} 440--448}	
	}
}	
\end{thebibliography}
\end{document}